\documentclass[12pt]{revtex4}
\usepackage{amsfonts}
\usepackage{graphicx, amsmath}
\usepackage{longtable}
\makeatletter

\newcommand{\Rmnum}[1]{\expandafter\@slowromancap\romannumeral #1@}
\makeatother
\begin{document}
\title[Short Title]{Fast generations of tree-type three-dimensional entanglement via Lewis-Riesenfeld invariants and transitionless quantum driving}
\author{Jin-Lei Wu}
\author{Xin Ji\footnote{E-mail: jixin@ybu.edu.cn}}
\author{Shou Zhang}
\affiliation{Department of Physics, College of Science, Yanbian University, Yanji, Jilin 133002, People's Republic of China}

\begin{abstract}
Recently, a novel three-dimensional entangled state called tree-type entanglement, which is likely to have applications for improving quantum communication security, was prepared via adiabatic passage by Song $et~al.$ [Phys. Rev. A \textbf{93}, 062321 (2016)]. Here we propose two schemes for fast generations of tree-type three-dimensional entanglement among three spatially separated atoms via shortcuts to adiabatic passage. With the help of quantum Zeno dynamics, two kinds of different but equivalent methods, Lewis-Riesenfeld invariants and transitionless quantum driving, are applied to construct shortcuts to adiabatic passage. The comparisons between the two methods are discussed. The strict numerical simulations show that the tree-type three-dimensional entangled states can be fast prepared with quite high fidelities and the two schemes are both robust against the variations in the parameters, atomic spontaneous emissions and the cavity-fiber photon leakages.
\\{\bf{Keywords:}} Three-dimensional entanglement, Quantum Zeno dynamics, Lewis-Riesenfeld invariants, Transitionless quantum driving
\end{abstract}
\maketitle
\section{Introduction}
Entanglement plays a crucial role in quantum information processing~\cite{AKE,CHBS,CD2000,GV}. Some of typical entangled states, such as Bell state~\cite{BJS1965}, Greenberger-Horne-Zeilinger~(GHZ) state~\cite{DMA1990} and W state~\cite{WGJ2000}, have attracted great attention in the last decades. However, all of these states are entangled states that are defined in Hilbert spaces with two dimensions. Recently, high-dimensional entanglement has attracted more and more attention due to their superior security than qubit systems, especially in the aspect of quantum key distribution. Besides, it has been demonstrated that violations of local realism by two entangled high-dimensional systems are stronger than that by two-dimensional systems~\cite{DPMW2000}. Thus, much interest has been focused on the generation of high-dimensional entanglement in theory via various techniques including quantum Zeno dynamics~(QZD)~\cite{WG2011,SJR2013, QCW2013}, stimulated Raman adiabatic passage~(STIRAP)~\cite{LP2012,YLS2015}, and dissipative dynamics~\cite{XQ201402,SL2014}. Also, experimental generations of high-dimensional entanglement have been already achieved~\cite{AAGA2001,AGA2002}.

With no doubt, a lot of remarkable achievements have been made with regard to high-dimensional entangled states. However, most of these high-dimensional entangled states are two-body but few multi-body. For a dozen years, some attention has been paid to multi-body high-dimensional entangled states such as singlet state~\cite{AC2002}, and lots of schemes have been proposed for generations of singlet state~\cite{XHL2010,ZYJ2013,MYJ2013,YYJ2014}. A short time before, a novel three-body three-dimensional entangled state called tree-type entanglement was prepared via adiabatic passage by Song $et~al.$~\cite{SSW2016}. In the reference~\cite{SSW2016}, the tree-type three-dimensional entanglement was prepared among one single atom and two BECs and the authors predicted that the tree-type three-dimensional entanglement is likely to have great applications in improving quantum communication security.

Among the techniques mentioned above for generations of high-dimensional entanglement, there are two techniques widely used for their robustness against decoherence in certain conditions. One is STIRAP~\cite{LP2012,MYJ2013,YLS2015,SSW2016}, and the other is QZD~\cite{WG2011,SJR2013, QCW2013,XHL2010,ZYJ2013,YYJ2014}. STIRAP is widely used in time-dependent interacting fields and robust against the atomic spontaneous emission and variations in the experimental parameters, but a relatively long interaction time is usually required. QZD is usually robust against photon leakages and does not need a long interaction time. However QZD is sensitive to the atomic spontaneous emissions and variations in the experimental parameters. Thus some of  researchers introduce detuning between the atomic transitions to restrain the influence of atomic spontaneous emissions~\cite{SJR2013}, but the interaction time significantly increases unavoidably. Therefore, in order to solve the problem of long interaction time, researchers have paid more attention to ``shortcut to adiabatic passage'' which employs a set of techniques to speed up a slow quantum adiabatic process~\cite{XASA2010,AC2013,YHC2014,YYQ2014,YLQ2015,YYJ2015,YLX2015,YLC2015,JYC2016,ZYY2016,YHC2016}, in which Chen $et~al.$ implemented entangled states' fast preparation via shortcut to adiabatic passage~\cite{YHC2014,YYQ2014,YYJ2015}, Lin $et~al.$ fast generated the two-atom three-dimensional entanglement via Lewis-Riesenfeld invariants~(LRI)~\cite{JYC2016}, and Chen $et~al.$ fast prepared the three-atom singlet state by transitionless quantum driving~(TQD)~\cite{ZYY2016}.

In this paper, we propose two schemes for fast generations of tree-type three-dimensional entanglement among three spatially separated atoms via LRI and TQD, respectively. Based on LRI and TQD we construct effective shortcuts to adiabatic passage for fast generating tree-type three-dimensional entanglement among three atoms trapped respectively in three spatially separated cavities connected by two fibers. We will give the interesting comparisons between the LRI method and the TQD method. The generations of tree-type three-dimensional entanglement in our schemes are implemented within a short time and the strict numerical simulations demonstrate that our schemes are both robust against the decoherence caused by the atomic spontaneous emissions, photon leakages and the variations in the parameters.

\section{Preliminary theory}
\label{sec2}

\subsection{Lewis-Riesenfeld invariants}
Here we give a brief description about Lewis-Riesenfeld invariants theory~\cite{HRL1969}. A quantum system is governed by a time-dependent Hamiltonian $H(t)$, and the corresponding time-dependent Hermitian invariant $I(t)$ satisfies
\begin{eqnarray}\label{e1}
i\hbar\frac{\partial I(t)}{\partial t}&=&[H(t),I(t)].
\end{eqnarray}
The solution of the time-dependent Schr\"odinger equation $i\hbar\partial_t|\Psi(t)\rangle=H(t)|\Psi(t)\rangle$ can be expressed by a superposition of invariant $I(t)$ dynamical modes $|\Phi_{n}(t)\rangle$
\begin{eqnarray}\label{e2}
|\Psi(t)\rangle&=&\sum_n C_n e^{i\alpha_n}|\Phi_{n}(t)\rangle,
\end{eqnarray}
where $C_n$ is the time-independent amplitude, $\alpha_n$ is the time-dependent Lewis-Riesenfeld phase, and $|\Phi_{n}(t)\rangle$ is one of the
orthogonal eigenvectors of the invariant $I(t)$ satisfying $I(t)|\Phi_{n}(t)\rangle=\lambda_n|\Phi_{n}(t)\rangle$ with a real eigenvalue $\lambda_n$. The Lewis-Riesenfeld phases
are defined as
\begin{eqnarray}\label{e3}
\alpha_n(t)&=&\frac{1}{\hbar}\int_0^t dt^\prime
\langle\Phi_n(t^\prime )|i\hbar\frac{\partial}{\partial t^\prime
}-H(t^\prime )|\Phi_n(t^\prime )\rangle.
\end{eqnarray}

\subsection{Transitionless quantum driving}
Suppose a system is dominated by a time-dependent Hamiltonian $H_0(t)$ with instantaneous eigenstates $|\phi_n(t)\rangle$ and eigenvalues $E_n(t)$,
\begin{eqnarray}\label{e4}
H_0(t)|\phi_n(t)\rangle=E_n(t)|\phi_n(t)\rangle.
\end{eqnarray}
When a slow change satisfying the adiabatic condition happens, the state of the system governed by $H_0(t)$ can be expressed as~\cite{XCE2011,MVB2009}
\begin{eqnarray}\label{e5}
|\psi(t)\rangle&=&e^{i\xi_n(t)}|\phi_n(t)\rangle,\nonumber\\
\xi_n(t)&=&-\frac{1}{\hbar}\int_0^t dt^\prime E_n(t^\prime )+i\int_0^t dt^\prime \langle\phi_n(t^\prime)|\partial_ {t^\prime} \phi_n(t^\prime)\rangle.
\end{eqnarray}
Because the instantaneous eigenstates $|\phi_n(t)\rangle$ do not meet the Schr\"{o}dinger equation, there may be transitions between the eigenstates of $H_0(t)$ with a finite probability during the whole evolution process even under the adiabatic condition. In order to construct the Hamiltonian $H(t)$ that exactly drives the instantaneous eigenstates $|\phi_n(t)\rangle$, i.e., there are no transitions between different eigenstates during the whole evolution process, the simplest choice of the Hamiltonian $H(t)$ can be written as
\begin{eqnarray}\label{e6}
H(t)=i\hbar\sum_n|\partial_ {t}\phi_n\rangle\langle\phi_n|.
\end{eqnarray}
Therefore, as long as $H(t)$ is constructed, the system will evolves with no transitions between different eigenstates.
\subsection{Quantum Zeno dynamics}
Assume that a quantum system's dynamics evolution is governed by the Hamiltonian
\begin{eqnarray}\label{e7}
H_K=H_{\rm obs}+K H_{\rm meas},
\end{eqnarray}
where $H_{\rm obs}$ can be viewed as the Hamiltonian of the quantum system investigated and $H_{\rm meas}$ as an additional interaction Hamiltonian performing the measurement. $K$ is a coupling constant, and in the strong coupling limit $K\rightarrow \infty$, the whole system is governed by the evolution operator~\cite{PGS2009}
\begin{eqnarray}\label{e8}
U(t)=\lim_{K\rightarrow \infty}{\rm exp}\Big[-it\sum_{n}(K\lambda_nP_n+P_nH_{\rm obs}P_n)\Big],
\end{eqnarray}
where $\sum_{n}P_nH_{obs}P_n$ is Zeno Hamiltonian, $P_n$ is one of the eigenprojections of $H_{\rm meas}$ with
eigenvalues $\lambda_n$($H_{\rm meas} = \sum_{n}\lambda_nP_n$). Interestingly, it is easy to deduce that the system state will remain in the same Zeno subspace as that of its initial state. In particular, if the system is initially in the dark state $|\Psi_d\rangle$ of $H_{\rm meas}$, i.e., $H_{\rm meas}|\Psi_d\rangle=0$, the evolution operator reduces to
\begin{eqnarray}\label{e9}
U(t)=\lim_{K\rightarrow \infty}{\rm exp}(-itP_nH_{\rm obs}P_n).
\end{eqnarray}

\section{Description of the physical model for generating tree-type three-dimensional entanglement}
\label{sec3}

\begin{figure}[htb]\centering
  % Requires \usepackage{graphicx}
  \includegraphics[scale=0.5]{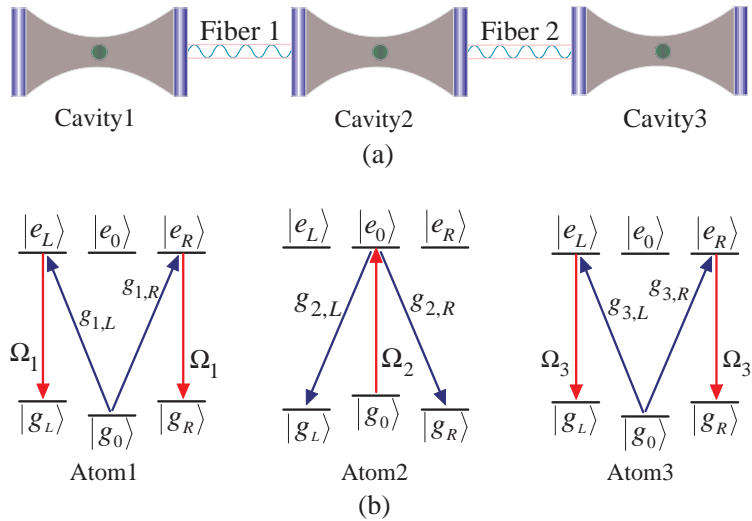}\\
  \caption{(Color online) (\textbf{a})~The schematic setup for generating tree-type three-dimensional entanglement; (\textbf{b})~the level configurations and relevant transitions. }\label{f1}
\end{figure}
The schematic setup for generating the tree-type three-dimensional entanglement is shown in Fig.~\ref{f1}(\textbf{a}). Three atoms are trapped respectively in three spatially separated optical cavities which are connected by two fibers. Under the short fiber limit $(lv)/(2\pi c)\leq 1$, only the resonant modes of the fibers interact with the cavity modes~\cite{SMB2006}, where $l$ is the length of the fiber and $v$ is the decay rate of the cavity field into a continuum of fiber modes. The atomic level configurations and relevant transitions are shown in Fig.~\ref{f1}\textbf{(b)}. As shown in Fig.~\ref{f1}\textbf{(b)}, the five-level atom1 and atom3 are both $M$-type with two excited states $|e_L\rangle$ and $|e_R\rangle$ and three ground states $|g_L\rangle$, $|g_0\rangle$ and $|g_R\rangle$. The four-level atom2 is tripod-type with one excited state $|e_0\rangle$ and three ground states $|g_L\rangle$, $|g_0\rangle$ and $|g_R\rangle$. The atomic transition $|e_{L(R)}\rangle_j\leftrightarrow|g_0\rangle_j$~$(j=1,3)$ is resonantly coupled to the left-circularly~(right-circularly) polarized mode of $j$th cavity with corresponding coupling constant $g_{j,L(R)}$, and $|e_0\rangle_2\leftrightarrow|g_{L(R)}\rangle_2$ is resonantly coupled to the left-circularly~(right-circularly) polarized mode of cavity2 with corresponding coupling constant $g_{2,L(R)})$. The transitions $|e_{L(R)}\rangle_j\leftrightarrow|g_{L(R)}\rangle_j$ and $|e_0\rangle_2\leftrightarrow|g_0\rangle_2$ are resonantly driven by classical laser fields with the time-dependent Rabi frequencies $\Omega_j(t)$ and $\Omega_2(t)$, respectively. Then the whole system can be dominated by the interaction Hamiltonian ($\hbar=1$):
\begin{eqnarray}\label{e10}
H_{\rm total}&=&H_{al}+H_{acf},\nonumber\\
H_{al}&=&\sum_{i=L,R}\sum_{j=1,3}\Omega_j(t)|e_i\rangle_j\langle{g_i}|+\Omega_2(t)(|e_0\rangle_2\langle{g_0}|+\rm H.c.,\nonumber\\
H_{acf}&=&\sum_{i=L,R}\Big[\sum_{j=1,3}g_{j,i}a_{j,i}|e_i\rangle_{j}\langle{g_0}|+g_{2,i}a_{2,i}|e_0\rangle_{2}\langle{g_i}|~\nonumber\\
&&\quad+\,v_{1,i}(a_{1,i}+a_{2,i})b_{1,i}^\dag+v_{2,i}(a_{2,i}+a_{3,i})b_{2,i}^\dag\Big]+\rm H.c.,
\end{eqnarray}
where $H_{\rm total}$ is the total Hamiltonian of the whole system, $H_{al}~(H_{acf})$ is the interaction between the atoms and the classical laser fields~(the cavity-fiber system), $v_{1(2),L(R)}$ is the coupling strength between the modes of the cavity1,2~(cavity2,3) and the modes of the fiber1(2), $a_{k,L(R)}~(k=1,2,3)$ is the annihilation operator of left-circularly~(right-circularly) polarized mode of $k$th cavity, and $b_{1(2),L(R)}^\dag$ is the creation operator of fiber1(2) left-circularly~(right-circularly) polarized mode. For simplicity, we assume $g_{k,L(R)}$ and $v_{1(2),L(R)}$ are real, $g_{k,L(R)}=g$, and $v_{1(2),L(R)}=v$.

Suppose that the total system is initially in the state $|\phi_1\rangle=|g_0\rangle_1|g_0\rangle_2|g_0\rangle_3|00000\rangle_{c_1f_1c_2f_2c_3}$ denoting $k$th atom in the state $|g_0\rangle_k$ and all of three cavities and two fibers in the vacuum state. Thus dominated by the total Hamiltonian in Eq.~(\ref{e10}), the whole system evolves in the Hilbert space spanned by
\begin{eqnarray}\label{e11}
    |\phi_1\rangle&=&|g_0\rangle_1|g_0\rangle_2|g_0\rangle_3|00000\rangle_{c_1f_1c_2f_2c_3},\quad\nonumber
    |\phi_2\rangle=|g_0\rangle_1|e_0\rangle_2|g_0\rangle_3|00000\rangle_{c_1f_1c_2f_2c_3},\nonumber\\
    |\phi_3\rangle&=&|g_0\rangle_1|g_L\rangle_2|g_0\rangle_3|00L00\rangle_{c_1f_1c_2f_2c_3},\quad\nonumber
    |\phi_4\rangle=|g_0\rangle_1|g_R\rangle_2|g_0\rangle_3|00R00\rangle_{c_1f_1c_2f_2c_3},\nonumber\\
    |\phi_5\rangle&=&|g_0\rangle_1|g_L\rangle_2|g_0\rangle_3|0L000\rangle_{c_1f_1c_2f_2c_3},\quad\nonumber
    |\phi_6\rangle=|g_0\rangle_1|g_L\rangle_2|g_0\rangle_3|000L0\rangle_{c_1f_1c_2f_2c_3},\nonumber\\
    |\phi_7\rangle&=&|g_0\rangle_1|g_R\rangle_2|g_0\rangle_3|0R000\rangle_{c_1f_1c_2f_2c_3},\quad\nonumber
    |\phi_8\rangle=|g_0\rangle_1|g_R\rangle_2|g_0\rangle_3|000R0\rangle_{c_1f_1c_2f_2c_3},\nonumber\\
    |\phi_9\rangle&=&|g_0\rangle_1|g_L\rangle_2|g_0\rangle_3|L0000\rangle_{c_1f_1c_2f_2c_3},\quad\nonumber
    |\phi_{10}\rangle=|g_0\rangle_1|g_L\rangle_2|g_0\rangle_3|0000L\rangle_{c_1f_1c_2f_2c_3},\nonumber\\
    |\phi_{11}\rangle&=&|g_0\rangle_1|g_R\rangle_2|g_0\rangle_3|R0000\rangle_{c_1f_1c_2f_2c_3}, \quad\nonumber
    |\phi_{12}\rangle=|g_0\rangle_1|g_R\rangle_2|g_0\rangle_3|0000R\rangle_{c_1f_1c_2f_2c_3}.\nonumber\\
    |\phi_{13}\rangle&=&|e_L\rangle_1|g_L\rangle_2|g_0\rangle_3|00000\rangle_{c_1f_1c_2f_2c_3}, \quad\nonumber
    |\phi_{14}\rangle=|g_0\rangle_1|g_L\rangle_2|e_L\rangle_3|00000\rangle_{c_1f_1c_2f_2c_3},\nonumber\\
    |\phi_{15}\rangle&=&|e_R\rangle_1|g_R\rangle_2|g_0\rangle_3|00000\rangle_{c_1f_1c_2f_2c_3}, \quad\nonumber
    |\phi_{16}\rangle=|g_0\rangle_1|g_R\rangle_2|e_R\rangle_3|00000\rangle_{c_1f_1c_2f_2c_3}.\nonumber\\
    |\phi_{17}\rangle&=&|g_L\rangle_1|g_L\rangle_2|g_0\rangle_3|00000\rangle_{c_1f_1c_2f_2c_3}, \quad\nonumber
    |\phi_{18}\rangle=|g_0\rangle_1|g_L\rangle_2|g_L\rangle_3|00000\rangle_{c_1f_1c_2f_2c_3},\nonumber\\
    |\phi_{19}\rangle&=&|g_R\rangle_1|g_R\rangle_2|g_0\rangle_3|00000\rangle_{c_1f_1c_2f_2c_3}, \quad
    |\phi_{20}\rangle=|g_0\rangle_1|g_R\rangle_2|g_R\rangle_3|00000\rangle_{c_1f_1c_2f_2c_3}.
\end{eqnarray}
Obviously, the system is initially in the dark state of $H_{acf}$, i.e., $H_{acf}|\phi_1\rangle=0$. Therefore, under the Zeno limit condition $\Omega_k(t) \ll g, v~(k=1,2,3)$, the whole system can approximatively evolve in an invariant Zeno subspace consisting of dark states corresponding to the zero eigenvalue of $H_{acf}$:
 \begin{eqnarray}\label{e12}
    H_P=\Big\{|\phi_1\rangle, |\Psi_D\rangle, |\phi_{17}\rangle, |\phi_{18}\rangle, |\phi_{19}\rangle, |\phi_{20}\rangle \Big\},
\end{eqnarray}
corresponding to the projections
\begin{equation}\label{e13}
    P^\alpha=|\alpha\rangle\langle\alpha|,\quad(|\alpha\rangle\in H_{P}).
\end{equation}
Here,
\begin{eqnarray}\label{e14}
    |\Psi_D\rangle&=&\frac{1}{2\sqrt{2v^2+g^2}}\Big[2v|\phi_2\rangle-g\Big(|\phi_5\rangle+|\phi_6\rangle+|\phi_7\rangle+|\phi_8\rangle\Big)\nonumber\\
    &&\quad+\,v\Big(|\phi_{13}\rangle+|\phi_{14}\rangle+|\phi_{15}\rangle+|\phi_{20}\rangle\Big)\Big].
\end{eqnarray}
Therefore, the system Hamiltonian can be rewritten as the following form based on Eq.~(\ref{e9}):
\begin{eqnarray}\label{e15}
H_{\rm total}&\simeq&\sum_{\alpha}P^\alpha H_{al}P^\alpha~\nonumber\\
&=&\frac{v}{\sqrt{2v^2+g^2}}\Big[\frac{1}{2}\Omega_1(t)\Big(|\phi_{17}\rangle+|\phi_{18}\rangle\Big)\langle\Psi_D|\nonumber\\
&&\quad+\,\Omega_2(t)|\phi_1\rangle\langle\Psi_D|+\frac{1}{2}\Omega_3(t)\Big(|\phi_{19}\rangle+|\phi_{20}\rangle\Big)\langle\Psi_D|\Big]+\rm H.c..
\end{eqnarray}
 Here setting $v=g$ and $\Omega_3(t)=\Omega_1(t)$, we can obtain an effective Hamiltonian of the system
\begin{eqnarray}\label{e16}
H_0(t)=\frac{1}{\sqrt3}\Big(\Omega_2(t)|\Psi_1\rangle+\Omega_1(t)|\Psi_2\rangle\Big)\langle\Psi_D|+\rm H.c..
\end{eqnarray}
in which $|\Psi_1\rangle=|\phi_1\rangle$, and $|\Psi_2\rangle=\frac{1}{2}(|\phi_{17}\rangle+|\phi_{18}\rangle+|\phi_{19}\rangle+|\phi_{20}\rangle)$. The instantaneous eigenstates of $H_0(t)$ corresponding respectively to the eigenvalues $\lambda_0=0$ and $\lambda_{\pm}=\pm \Omega(t)/\sqrt{3}$ are
\begin{eqnarray}\label{e17}
|n_0(t)\rangle=\left(
\begin{array}{c}
-\cos\theta(t)\\
0\\
\sin\theta(t)
\end{array}\right),\quad
|n_\pm(t)\rangle=\frac{1}{\sqrt{2}}\left(
\begin{array}{c}
\sin\theta(t)\\
\pm 1\\
\cos\theta(t)
\end{array}\right),
\end{eqnarray}
where $\Omega(t)=\sqrt{\Omega_1(t)^2+\Omega_2^2(t)}$ and $\tan\theta(t)=\Omega_2(t)/\Omega_1(t)$.

\section{Two methods used to generate tree-type three-dimensional entanglement}
\label{sec4}
\subsection{The method of Lewis-Riesenfeld invariants}

In order to construct a shortcut by the LRI method for fast generation of tree-type three-dimensional entanglement, we are supposed to chase down the Hermitian invariant operator $I(t)$ satisfying $i\hbar \partial_t I(t)=[H_{0}(t),I(t)]$. Because of SU(2) dynamical symmetry of $H_{0}(t)$ in Eq.~(\ref{e16}), $I(t)$ can be easily given by~\cite{XCE2011}
\begin{eqnarray}\label{e18}
I(t)&=&=\frac{1}{\sqrt3}\chi\left(
\begin{array}{ccc}
0                                        & \cos\nu\sin\beta                   & -i\sin\nu    \\
\cos\nu\sin\beta                         & 0                                  & \cos\nu\cos\beta \\
i\sin\nu                                 & \cos\nu\cos\beta                   & 0
\end{array}
\right).
\end{eqnarray}
where $\chi$ is an arbitrary constant in the unit of frequency keeping $I(t)$ in the unit of energy, and $\nu$ and $\beta$ are time-dependent auxiliary parameters satisfying the equations
\begin{eqnarray}\label{e19}
\dot{\nu}&=&\frac{1}{\sqrt3}(\Omega_{2}(t)\cos\beta-\Omega_{1}(t)\sin\beta),\nonumber\\
\dot{\beta}&=&\frac{1}{\sqrt3}\tan\nu(\Omega_{1}(t)\cos\beta+\Omega_{2}(t)\sin\beta).
\end{eqnarray}
So $\Omega_{1}(t)$ and $\Omega_{2}(t)$ can be easily deduced as follows:
\begin{eqnarray}\label{e20}
\Omega_{1}(t)&=&\sqrt3(\dot{\beta}\cot\nu\cos\beta-\dot{\nu}\sin\beta),\nonumber\\
\Omega_{2}(t)&=&\sqrt3(\dot{\beta}\cot\nu\sin\beta+\dot{\nu}\cos\beta).
\end{eqnarray}
The solution of Shr\"{o}dinger equation $i\hbar\partial_t|\Psi(t)\rangle=H_{0}(t)|\Psi(t)\rangle$ with respect to the instantaneous eigenstates of $I(t)$ can be written as $|\Psi(t)\rangle=\sum_{n=0,\pm}C_ne^{i\alpha_n}|\phi_n(t)\rangle$, where $\alpha_n(t)$ is the Lewis-Riesenfeld phase in Eq.~(\ref{e3}), $C_n=\langle\phi_n(0)|\Psi_1\rangle$, and $|\phi_n(t)\rangle$ is the eigenstate of the invariant $I(t)$ as
\begin{eqnarray}\label{e21}
|\phi_0(t)\rangle=\left(
\begin{array}{c}
\cos\nu\cos\beta\\
-i\sin\nu\\
-\cos\nu\sin\beta
\end{array}\right), \quad
|\phi_\pm(t)\rangle=\frac{1}{\sqrt{2}}\left(
\begin{array}{c}
\sin\nu\cos\beta\pm i\sin\beta\\
i\cos\nu\\
-\sin\nu\sin\beta\pm i\cos\beta
\end{array}\right).
\end{eqnarray}
Then we consider a series of boundary conditions satisfying $[H_0(0), I(0)]=0$ and $[H_0(t_f), I(t_f)]=0$ to give
\begin{eqnarray}\label{e22}
\lim_{t\to 0}\frac{\Omega_2(t)}{\Omega_1(t)}=0,\quad\lim_{t\to t_f}\frac{\Omega_2(t)}{\Omega_1(t)}=2.
\end{eqnarray}
where $t_f$ is the operation time. The Eq.~(\ref{e22}) is the guarantee for the system to evolve along $|n_0(t)\rangle$ in Eq.~(\ref{e17}) so that we obtain the target state $|\Psi_{LRI}\rangle=\frac{1}{\sqrt5}|\Psi_{1}\rangle-\frac{2}{\sqrt5}|\Psi_{2}\rangle$.
Therefore, to avoid infinite Rabi frequencies, we can choose the boundary conditions for $\nu$ and $\beta$ as follows:
\begin{eqnarray}\label{e23}
\nu(0)&=&\varepsilon,\quad\dot{\nu}(0)=0,\quad\nu(t_f)=\varepsilon,\quad\dot{\nu}(t_f)=0,\nonumber\\
~\beta(0)&=&0,\quad\beta(t_f)=\arctan2.
\end{eqnarray}
where $\varepsilon$ is a time-independent small value. Then the parameters can be easily set as
\begin{eqnarray}\label{e24}
\nu(t)=\varepsilon,\quad\beta(t)=\frac{\arctan2~t}{t_f},
\end{eqnarray}
and thus we can deduce
\begin{eqnarray}\label{e25}
\Omega_{1}(t)&=&\frac{\sqrt3\arctan2}{t_f}\cot\varepsilon\cos\frac{\arctan2~t}{t_f},\nonumber\\
\Omega_{2}(t)&=&\frac{\sqrt3\arctan2}{t_f}\cot\varepsilon\sin\frac{\arctan2~t}{t_f}.
\end{eqnarray}
Based on the parameters above, we can determine the value of $\varepsilon$ by calculating the fidelity
\begin{eqnarray}\label{e26}
F&=&|\langle\Psi_{LRI}|\Psi(t_f)\rangle|^2,\nonumber\\
&=&\Big[\cos^2\varepsilon+\sin^2\varepsilon~\cos\Big(\frac{\arctan2}{\sin\varepsilon}\Big)\Big]^2,
\end{eqnarray}
where $|\Psi(t_f)\rangle=\sum_{n=0,\pm}C_ne^{i\alpha_n(t_f)}|\phi_n(t_f)\rangle$ with the Lewis-Riesenfeld phases
\begin{eqnarray}\label{e27}
\alpha_0(t_f)=0,\quad
\alpha_{\pm}(t_f)=\mp \frac{\arctan2}{\sin\varepsilon},
\end{eqnarray}
where $|\Psi_{LRI}\rangle=\frac{1}{\sqrt5}\Big(|\phi_{1}\rangle-|\phi_{17}\rangle-|\phi_{18}\rangle-|\phi_{19}\rangle-|\phi_{20}\rangle\Big)$ is the tree-type three-dimensional entanglement generated by the LRI method.
Therefore, for the appropriate Rabi frequencies and the fidelity $F=1$, we can choose
\begin{eqnarray}\label{e28}
\frac{\arctan2}{\sin\varepsilon}=2\pi,\quad\rm i.e.\quad
\varepsilon=\arcsin\left(\frac{\arctan2}{2\pi}\right)=0.177.
\end{eqnarray}
Thus, the transformation $|\Psi_1\rangle\rightarrow|\Psi_{LRI}\rangle$ is achieved and we have constructed a shortcut by the LRI method to speed up the generation of the tree-type three-dimensional entanglement.
\subsection{The method of transitionless quantum driving}
\begin{figure}[htb]\centering
 % Requires \usepackage{graphicx}
  \includegraphics[scale=0.55]{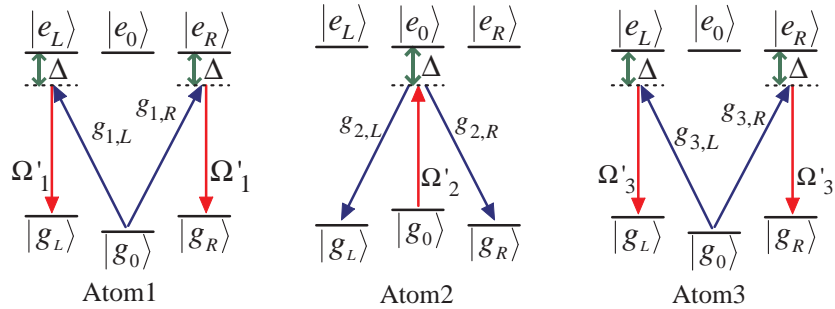}\\
  \caption{(Color online) The APF design of the schematic setup for TQD to fast generate tree-type three-dimensional entanglement}\label{f2}
\end{figure}
Because the instantaneous eigenstates of $H_0(t)$ do not meet the Schr\"{o}dinger equation, there may be transitions between the eigenstates in Eq.~(\ref{e17}). Therefore, we need to construct the TQD Hamiltonian $H(t)$ that exactly drives the instantaneous eigenstates with no transitions between different eigenstates. Based on Eq.~(\ref{e6}), we learn the
simplest Hamiltonian $H(t)$ is derived in the form
\begin{eqnarray}\label{e29}
H(t)=i\sum_{k=0,\pm}|\partial_ {t}n_k(t)\rangle\langle n_k(t)|.
\end{eqnarray}
Substituting Eq.~(\ref{e17}) into Eq.~(\ref{e29}), we obtain
\begin{eqnarray}\label{e30}
H(t)=i\dot{\theta}|\Psi_1\rangle\langle \Psi_2|+\rm H.c.,
\end{eqnarray}
in which $\dot{\theta}(t)=\Big(\dot\Omega_2(t)\Omega_1(t)-\Omega_2(t)\dot\Omega_1(t)\Big)\Big/\Omega(t)$. According to our present system, the Hamiltonian $H(t)$ is too hard to be implemented under current experimental conditions. Fortunately, inspired by the references~\cite{YYJ2015,ZYY2016}, we find an alternative physically feasible~(APF) Hamiltonian whose effect is equivalent to $H(t)$. The APF design is shown in Fig.~\ref{f2}. Comparing Fig.~\ref{f1} and Fig.~\ref{f2}, we change all of the resonant atomic transitions into non-resonant atomic transitions with detuning $\Delta$.

The interaction Hamiltonian of the non-resonant system reads
\begin{eqnarray}\label{e31}
H'_{\rm total}&=&H'_{al}+H'_{acf}+H_e,\nonumber\\
H'_{al}&=&\sum_{i=L,R}\sum_{j=1,3}\Omega'_j(t)|e_i\rangle_j\langle{g_i}|+\Omega'_2(t)(|e_0\rangle_2\langle{g_0}|+\rm H.c.,\nonumber\\
H'_{acf}&=&\sum_{i=L,R}\Big[\sum_{j=1,3}g_{j,i}a_{j,i}|e_i\rangle_{j}\langle{g_0}|+g_{2,i}a_{2,i}|e_0\rangle_{2}\langle{g_i}|~\nonumber\\
&&+v_{1,i}(a_{1,i}+a_{2,i})b_{1,i}^\dag+v_{2,i}(a_{2,i}+a_{3,i})b_{2,i}^\dag\Big]+\rm H.c.,\nonumber\\
H_e&=&\sum_{i=L,R}\sum_{j=1,3}\Delta|e_i\rangle_j\langle e_i|+\Delta|e_0\rangle_2\langle e_0|.
\end{eqnarray}
Then similar to the approximation for the Hamiltonian from Eq.~(\ref{e10}) to Eq.~(\ref{e16}), an effective Hamiltonian for the present non-resonant system can be obtained
\begin{eqnarray}\label{e32}
H'_0(t)=\frac{1}{\sqrt3}\Big[|\Psi_D\rangle\Big(\Omega'_2(t)\langle\Psi_1|+\Omega'_1(t)\langle\Psi_2|\Big)+\rm H.c.\Big]+\Delta|\Psi_D\rangle\langle\Psi_D|.
\end{eqnarray}
Under the limit condition $\Delta\gg\Omega'_1(t)/\sqrt3, \Omega'_2(t)/\sqrt3$, by adiabatically eliminating the state $|\Psi_D\rangle$, the effective Hamiltonian $H'_0(t)$ becomes
\begin{eqnarray}\label{e33}
H_{eff}(t)=\frac{1}{3\Delta}\Big[|\Omega'_2(t)|^2|\Psi_1\rangle\langle\Psi_1|+|\Omega'_1(t)|^2|\Psi_2\rangle\langle\Psi_2|
+\Big(\Omega'^\ast_1(t)\Omega'_2(t)|\Psi_2\rangle\langle\Psi_1|+\rm H.c.\Big)\Big].
\end{eqnarray}
The first two terms can be removed by setting $\Omega'_1(t)=i~\Omega'_2(t)$ and the final effective Hamiltonian becomes
\begin{eqnarray}\label{e34}
H'_{eff}(t)=i~\frac{\Omega'_2(t)^2}{3\Delta}|\Psi_1\rangle\langle\Psi_2|+\rm H.c..
\end{eqnarray}
This effective Hamiltonian is equivalent to the Hamiltonian $H(t)$ in
Eq.~(\ref{e30}) if we set $\dot{\theta}(t)=\Omega'_2(t)^2/3\Delta$, i.e.,
\begin{eqnarray}\label{e35}
\Omega'_2(t)=\sqrt{\frac{3\Delta\Big(\dot\Omega_2(t)\Omega_1(t)-\Omega_2(t)\dot\Omega_1(t)\Big)}{\Omega(t)}}
\end{eqnarray}
which is the correlation between the Rabi frequencies of the TQD method and the Rabi frequencies of STIRAP. By setting the Rabi frequencies of STIRAP to satisfy the same boundary conditions as Eq.~(\ref{e22}), we can achieve the transformation $|\Psi_1\rangle\rightarrow|\Psi_{TQD}\rangle$ to implement the fast generation of tree-type three-dimensional entanglement, where $|\Psi_{TQD}\rangle=\frac{1}{\sqrt5}\Big(|\phi_{1}\rangle+|\phi_{17}\rangle+|\phi_{18}\rangle+|\phi_{19}\rangle+|\phi_{20}\rangle\Big)$ is the tree-type three-dimensional entanglement generated by the TQD method.

\section{Numerical simulations and comparisons between LRI and TQD}
\label{sec5}

In the following, we will give the numerical simulations in three subsections to discuss respectively the selections of parameters of the two methods, the feasibility of generating tree-type three-dimensional entanglement and the robustness of our schemes. Also the comparisons between the LRI method and the TQD method will be included in every subsection.
\subsection{Selections of parameters}\label{sub1sec5}
\begin{figure}[htb]\centering
 % Requires \usepackage{graphicx}
  \includegraphics[scale=0.75]{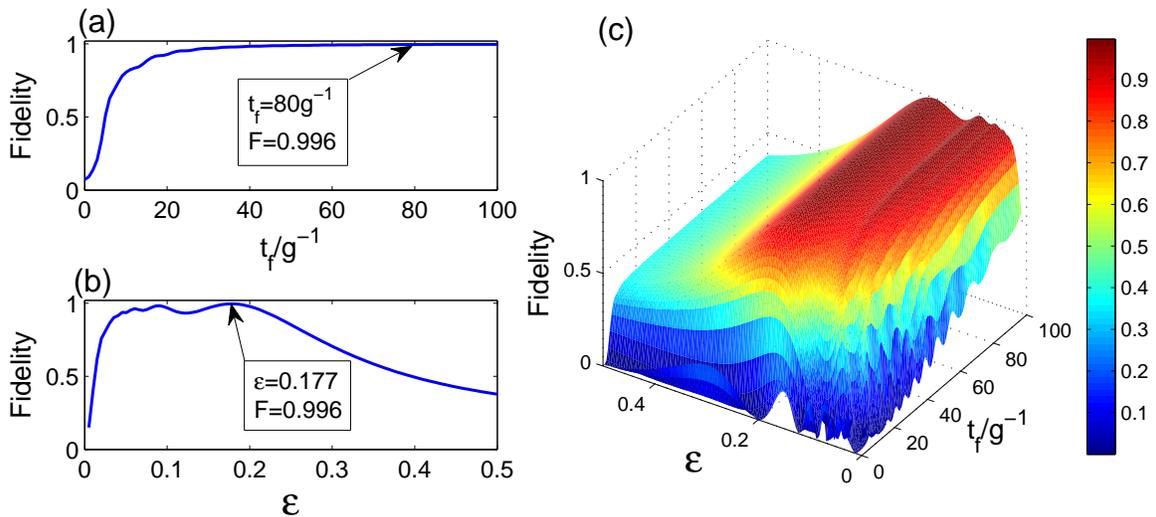}\\
\caption{(Color online) The fidelity for the LRI method versus \textbf{(a)} $t_f/g^{-1}$ with $\varepsilon=0.177$ and \textbf{(b)} $\varepsilon$ with $t_f=80/g$, respectively; \textbf{(c)}~the three dimensional image of the fidelity for the LRI method versus $t_f/g^{-1}$ and $\varepsilon$.}\label{f3}
\end{figure}
Firstly, to determine the parameters of the LRI method, we plot the fidelity $F=|\langle\Psi_{LRI}|\Phi(t_f)\rangle|^2$ versus the operation time $t_f$ and $\varepsilon$ in Fig.~\ref{f3}, where $|\Phi(t_f)\rangle$ is the state at the time $t=tf$ of the whole system governed by the total Hamiltonian $H_{\rm total}$ in Eq.~(\ref{e10}). In Fig.~\ref{f3}\textbf{(a)}, we plot the relation between the fidelity and the operation time $t_f$ with $\varepsilon=0.177$ which is determined in Eq.~(\ref{e28}). And we can see that in a very short operation time $t_f=80/g$ the fidelity is already almost unity: $F(t_f=80/g)=0.996$. From Fig.~\ref{f3}\textbf{(b)}, we can find that under $t_f=80/g$ when $\varepsilon=0.177$ the fidelity is highest. Thus we can choose $t_f=80/g$ and $\varepsilon=0.177$ as the parameters of the LRI method in the following discussion. Furthermore, in order to consider the joint effects of $t_f$ and $\varepsilon$ on the fidelity we plot the three dimensional image of the fidelity versus $t_f/g^{-1}$ and $\varepsilon$ in Fig.~\ref{f3}\textbf{(c)}. From the three dimensional image, it is clear that the effects of $t_f$ and $\varepsilon$ on the fidelity are not dependent on each other.

\begin{figure}[htb]\centering
  % Requires \usepackage{graphicx}
  \includegraphics[scale=0.55]{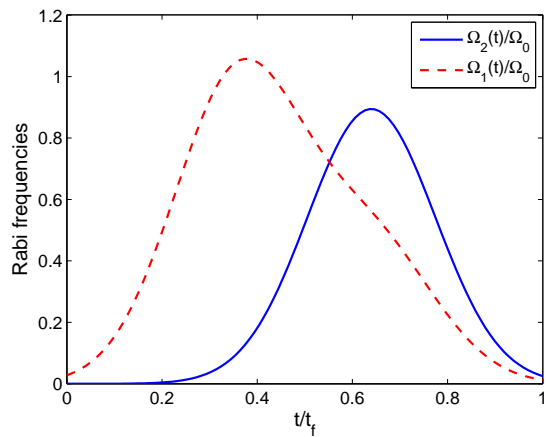}\\
  \caption{(Color online) Time dependence on $t/t_f$ of $\Omega_1(t)$ and $\Omega_2(t)$ with the parameters $\tau=0.14t_f$, $T=0.19t_f$.}\label{f4}
\end{figure}

Next we determine the parameters of the TQD method. In order to satisfy the boundary conditions in Eq.~(\ref{e22}), the Rabi frequencies $\Omega_1(t)$ and $\Omega_2(t)$ in the original Hamiltonian $H_{\rm total}$ are chosen as~\cite{ZYY2016}
\begin{eqnarray}\label{e36}
\Omega_1(t)&=&\frac{1}{\sqrt5}\Omega_0e^{-(t-t_f/2-\tau)^2/T^2}+\Omega_0e^{-(t-t_f/2+\tau)^2/T^2},\nonumber\\
\Omega_2(t)&=&\frac{2}{\sqrt5}\Omega_0e^{-(t-t_f/2-\tau)^2/T^2},
\end{eqnarray}
 where $\Omega_0$ is the pulses' amplitude, $t_f$ is the operation time, and $\tau$ and $T$ are related parameters. The time-dependent $\Omega_1(t)$ and $\Omega_2(t)$ are shown in Fig.~\ref{f4}.
\begin{figure}[htb]\centering
 % Requires \usepackage{graphicx}
  \includegraphics[scale=0.75]{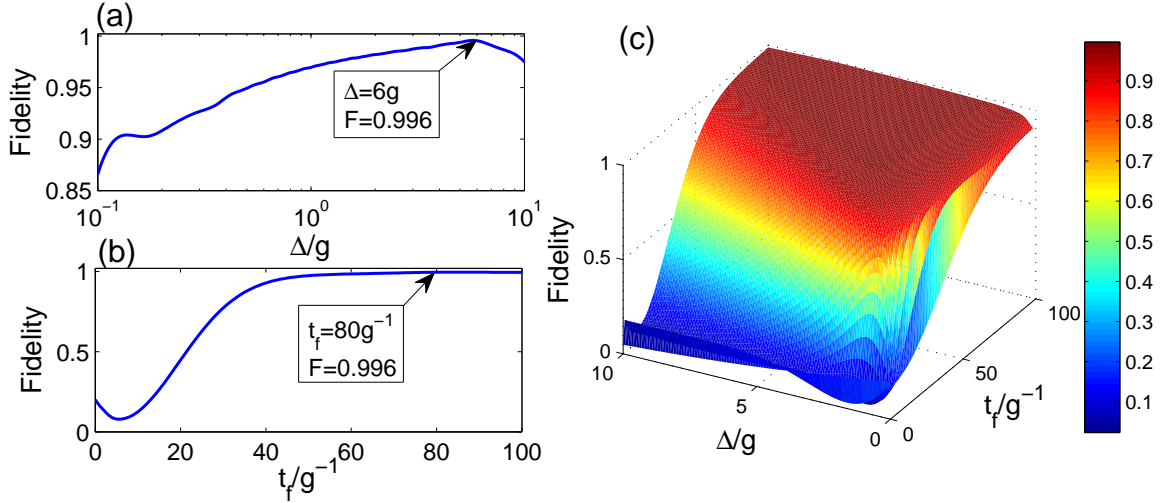}\\
\caption{(Color online) The fidelity for the TQD method versus \textbf{(a)} $\Delta/g$ with $t_f=80/g$ and \textbf{(b)} $t_f/g^{-1}$ with $\Delta=6g$, respectively; \textbf{(c)}~the three dimensional image of the fidelity for the TQD method versus $\Delta/g$ and $t_f/g^{-1}$.}\label{f5}
\end{figure}

Based on the correlation in Eq.~(\ref{e35}), the Rabi frequencies of the TQD method can be figured out. As an illustration, we plot the fidelity $F=|\langle\Psi_{TQD}|\Phi(t_f)\rangle|^2$ versus the detuning $\Delta$ and $t_f$ in Fig.~\ref{f5}, where $|\Phi(t_f)\rangle$ is the state at the time $t=t_f$ of the whole system governed by the total Hamiltonian $H'_{\rm total}$ in Eq.~(\ref{e31}). To compare with each other effectively, we choose the same operation time $t_f=80/g$ in the TQD method as that in the LRI method. From Fig.~\ref{f5}\textbf{(a)}, we can find that under $t_f=80/g$ when $\Delta=6g$ the fidelity is highest. Besides, we can see that the fidelity is almost unity: $F(t_f=80/g)=0.996$ at the point $t_f=80/g$ from Fig.~\ref{f5}\textbf{(b)}. Thus we choose $t_f=80/g$ and $\Delta=6g$ as the parameters of the TQD method in the following discussion. Similar to the LRI method, in order to consider the joint effects of $t_f$ and $\Delta$ on the fidelity we plot the three dimensional image of the fidelity versus $t_f/g^{-1}$ and $\Delta/g$ in Fig.~\ref{f5}\textbf{(c)}. However, from Fig.~\ref{f5}\textbf{(c)}, we are not able to judge whether the effects of $t_f$ and $\Delta$ on the fidelity are dependent or not dependent on each other. We will make a detailed discussion about the joint effects of $t_f$ and $\Delta$ on the fidelity later in the Subsection.
\subsection{Discussion of feasibility}\label{sub2sec5}
\begin{figure}[htb]\centering
  % Requires \usepackage{graphicx}
  \includegraphics[scale=0.75]{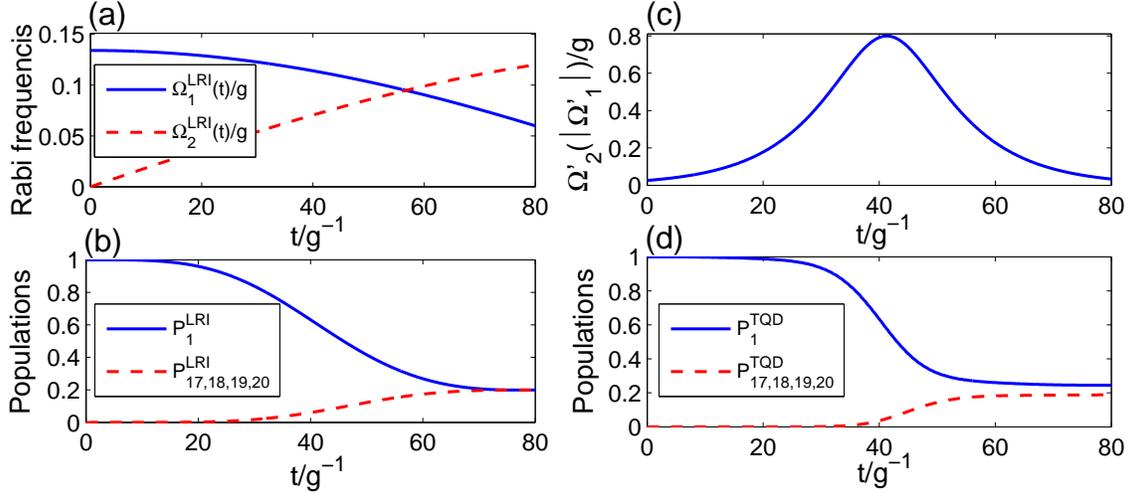}\\
\caption{(Color online) For the LRI method, \textbf{(a)}~the time dependence of the Rabi frequencies $\Omega_1(t)$~(blue solid line) and $\Omega_2(t)$~(red dashed line); \textbf{(b)}~the populations of states $|\phi_1\rangle$ and $|\phi_{17\sim20}\rangle$ governed by $H_{\rm total}$. For the TQD method, \textbf{(c)}~the time dependence of the Rabi frequency $\Omega'_2(t)~(|\Omega'_1(t)|)$; \textbf{(d)}~the populations of states $|\phi_1\rangle$ and $|\phi_{17\sim20}\rangle$ governed by $H'_{\rm total}$. The parameters used here are $t_f=80/g$, $\varepsilon=0.177$, $\Delta=6g$, $\tau=0.14t_f$ and $T=0.19t_f$.}\label{f6}
\end{figure}
\begin{figure}[htb]\centering
  % Requires \usepackage{graphicx}
  \includegraphics[scale=0.75]{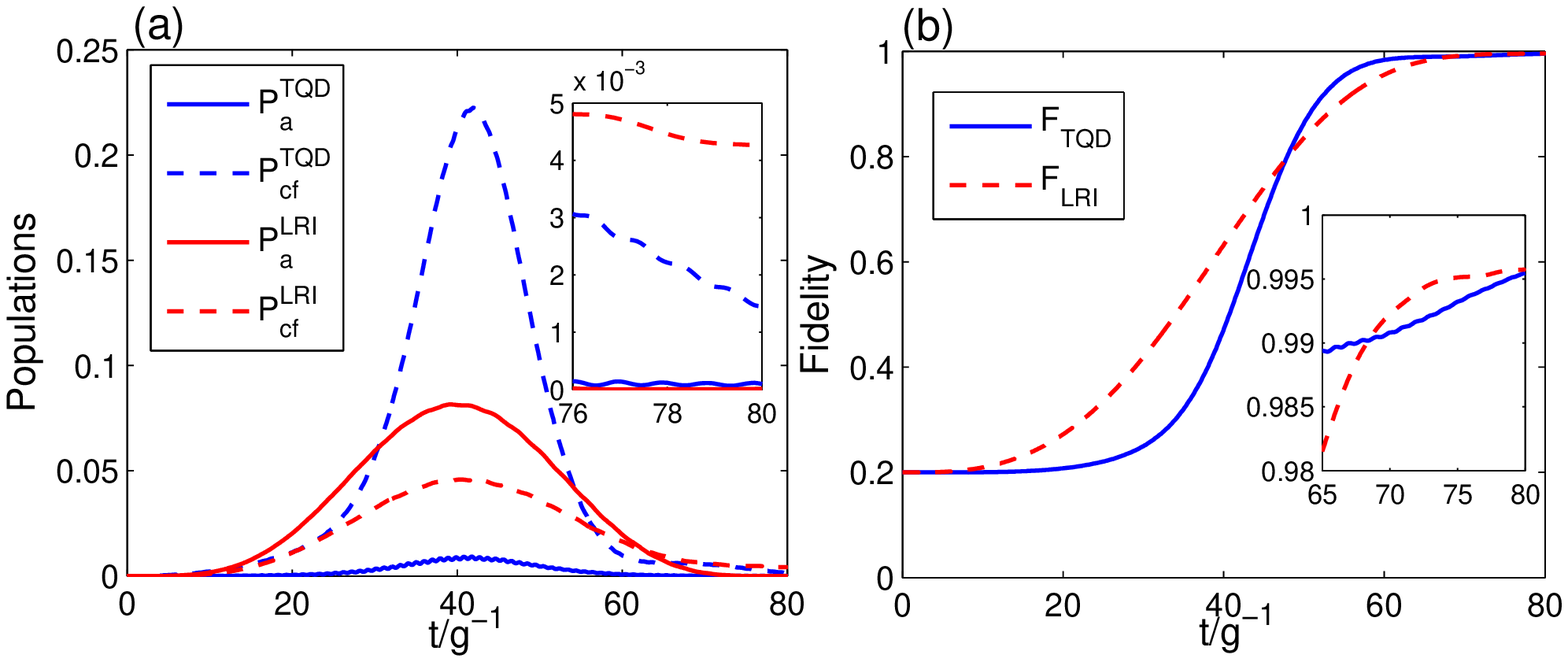}\\
\caption{(Color online) \textbf{(a)}~The time dependence of atomic excited populations~(red solid line) and cavity-fiber excited populations~(red dashed line) in the LRI method, and the time dependence of atomic excited populations~(blue solid line) and cavity-fiber excited populations~(blue dashed line) in the TQD method; \textbf{(b)}~the fidelities of tree-type three-dimensional entanglement for the LRI method~(red dashed line) and the TQD method~(blue solid line). The parameters used here are same as in Fig.~\ref{f6}.}\label{f7}
\end{figure}
In this subsection, we will give the numerical simulations for discussing the feasibility of our two schemes. For the LRI method, we plot the time-dependent Rabi frequencies $\Omega_1(t)$ and $\Omega_2(t)$ which are described by Eq.~(\ref{e25}) in Fig.~\ref{f6}\textbf{(a)}. The evolutions of populations $P_{1(17\sim20)}^{LRI}$ of states $|\phi_{1(17\sim20)}\rangle$ governed by $H_{\rm total}$ are shown in Fig.~\ref{f6}\textbf{(b)}. For the TQD method, we plot the time-dependent Rabi frequency $\Omega'_2(t)~(|\Omega'_1(t)|)$ which is described by Eq.~(\ref{e35}) in Fig.~\ref{f6}\textbf{(c)}. The evolutions of populations $P_{1(17\sim20)}^{TQD}$ of states $|\phi_{1(17\sim20)}\rangle$ governed by $H'_{\rm total}$ are shown in Fig.~\ref{f6}\textbf{(d)}. In addition, in Fig.~\ref{f7}\textbf{(a)} we plot the atomic excited populations $P_a^{LRI}=\sum_{m=2,13\sim16}P_m^{LRI}$~(red solid line) and $P_a^{TQD}=\sum_{m=2,13\sim16}P_m^{TQD}$~(blue solid line) corresponding to the LRI method and the TQD method respectively, and the cavity-fiber excited populations $P_{cf}^{LRI}=\sum_{n=3\sim12}P_{n}^{LRI}$~(red dashed line) and $P_{cf}^{TQD}=\sum_{n=3\sim12}P_{n}^{TQD}$~(blue dashed line) corresponding to the cases of the LRI method and the TQD method respectively. In Fig.~\ref{f7}\textbf{(b)}, we plot the fidelities of the tree-type three-dimensional entanglement generated by the LRI method~(red dashed line) and the TQD method~(blue solid line), respectively.

Here we first consider Fig.~\ref{f7}\textbf{(b)}. From Fig.~\ref{f7}\textbf{(b)}, we know that both of the two lines of fidelity illustrated based on the two methods reach a very high point close to unity at the time $t=80/g$, and thus both of our schemes are feasible. Fig.~\ref{f7}\textbf{(b)} also shows that the fidelity of the TQD method can reach a very high value within a shorter time than that of the LRI method. Next we consider the populations of the target states in Fig.~\ref{f6}\textbf{(b)} and Fig.~\ref{f6}\textbf{(d)}. We can see from Fig.~\ref{f6}\textbf{(b)} that a near perfect result we expect is obtained, but a little bit imperfect result appears in Fig.~\ref{f6}\textbf{(d)} in which there exists a small gap between two lines of the populations of $|\phi_1\rangle$ and $|\phi_{17\sim20}\rangle$. Therefore, for the transformation of populations, the LRI method is a bit better than the TQD method. However, when compare the pulse types in Fig.~\ref{f6}\textbf{(a)} and Fig.~\ref{f6}\textbf{(c)}, we find that the TQD method is more feasible than the LRI method. Because the pulses in LRI method are short-time truncations of two harmonic pulses and the truncations of the two harmonic pulses in a short time are too hard to be achieved. But the pulses in TQD method are almost complete Gaussian pulses which are relatively easy to be achieved. Moreover, the populations of atomic and cavity-fiber excited states for two methods are shown in Fig.~\ref{f7}\textbf{(a)} and all of the populations of excited states are near zero at the time $t=80/g$. So we can deduce that whichever method employed, the state of the whole system almost populates in tree-type three-dimensional entanglement.

It is worth explaining the gap between the two lines of the populations of $|\phi_1\rangle$ and $|\phi_{17\sim20}\rangle$ in Fig.~\ref{f6}\textbf{(d)}. For the TQD method, there are two limit conditions $\Omega'_k(t) \ll g, v~(k=1,2,3)$ and $\Delta \gg g, v$ applied to prepare tree-type three-dimensional entanglement. However, as shown in Fig.~\ref{f6}\textbf{(c)}, the amplitude of $\Omega'_2(t)~(|\Omega'_1(t)|)$ is $0.8g$ which do not strictly meet the limit condition $\Omega'_k(t) \ll g, v$. And also the detuning $\Delta=6g$ do not strictly meet the limit condition $\Delta \gg g, v$. In fact, these two limit conditions are difficult to be coordinated. With no assignments of $t_f$ and $\Delta$, we calculate the amplitude of $\Omega'_2(t)$ with the parameters $\tau=0.14t_f$ and $T=0.19t_f$
\begin{eqnarray}\label{e37}
\Omega'_0\approx\Omega'_2(t)\Big|_{t=\frac{1}{2}t_f}=2.9\sqrt{\frac{\Delta}{t_f}},
\end{eqnarray}
which is not dependent on the amplitude $\Omega_0$ of $\Omega_{1,2}(t)$ in Eq.~(\ref{e36}) but only proportional to $\sqrt{\Delta/t_f}$. Thus, $\Omega'_0$ roughly equals to $\sqrt{\Delta}/3$ if the operation time is chosen as $t_f=80/g$. Nevertheless, the ratio $1/3$ is not small enough to satisfy both two limit conditions $\Omega'_k(t) \ll g, v$ and $\Delta \gg g, v$, i.e., the condition $\Omega'_k(t) \ll g, v$ will not be satisfied if the limit condition $\Delta \gg g, v$ is satisfied and vice versa. Therefore, there exists a gap between the two lines of the populations of $|\phi_1\rangle$ and $|\phi_{17\sim20}\rangle$ in Fig.~\ref{f6}\textbf{(d)}. In addition, Eq.~(\ref{e37}) reveals that $\Omega'_2(t)$'s amplitude $\Omega'_0\propto\sqrt{\Delta/t_f}$. It is known that the fidelity of the TQD method is strongly dependent on $\Omega'_0$, so we can deduce that the fidelity of the TQD method is strongly dependent on the value of $\Delta/t_f$. As a result, in Fig.~\ref{f5}\textbf{(c)}, the effects of $t_f$ and $\Delta$ on the fidelity are dependent on each other.

Based on the discussion above, for fast generating tree-type three-dimensional entanglement,  both the LRI method and the TQD method are feasible. Besides, the two methods have their own advantages and disadvantages and we can choose a certain method depending on the conditions in experiment.
\subsection{Discussion of robustness}\label{sub3sec5}
\begin{figure}[htb]\centering
  % Requires \usepackage{graphicx}
  \includegraphics[scale=0.75]{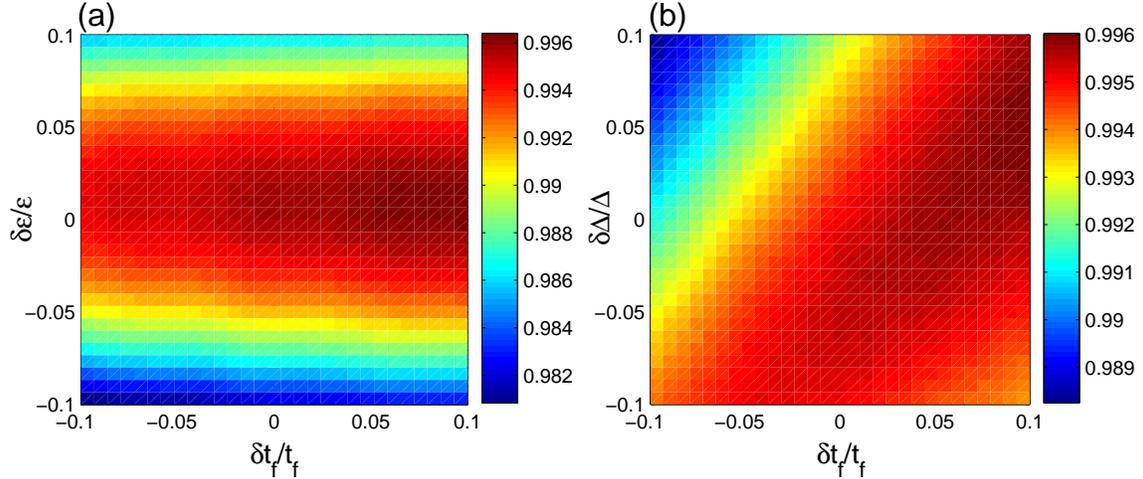}\\
\caption{(Color online) \textbf{(a)}~The fidelity of the LRI method versus $\delta t_f/t_f$ and $\delta\varepsilon/\varepsilon$; \textbf{(b)}~the fidelity of the TQD method versus $\delta t_f/t_f$ and $\delta\Delta/\Delta$. The parameters used here are same as in Fig.~\ref{f6}.}\label{f8}
\end{figure}
In the above discussion, the operations and the whole system are perfect and considered as absolutely isolated from the environment. Therefore, it is necessary to give the discussions of robustness of our schemes against the variations in the parameters and decoherence induced by the atomic spontaneous emissions and photon leakages of the cavity-fiber system. For discussing the effects of the variations in the parameters, we plot the fidelity of the LRI method versus the variations in $t_f$ and $\varepsilon$ in Fig.~\ref{f8}~\textbf{(a)} and the fidelity of the TQD method versus the variations in $t_f$ and $\Delta$ in Fig.~\ref{f8}~\textbf{(b)}. Here we define $\delta x= x'-x$ as the deviation of $x$, in which $x$ denotes the ideal value and $x'$ denotes the actual value. In Fig.~\ref{f8}~\textbf{(a)}, the fidelity decreases with the increase of $|\delta\varepsilon|$ as described in Fig.~\ref{f3}~\textbf{(b)}. From Eq.~(\ref{e25}), we know that the Rabi frequencies decrease with the increase of the operation time $t_f$. According to the limit condition $\Omega_k(t)\ll g, v$ we use, the values of the Rabi frequencies are the smaller the better, so the operation time $t_f$ is the longer the better as described in Fig.~\ref{f3}~\textbf{(a)}. Therefore, the fidelity of the LRI method increases with the increase of $\delta t_f$ in Fig.~\ref{f8}~\textbf{(a)}. In Fig.~\ref{f8}~\textbf{(b)}, we can clearly see that the effects of $t_f$ and $\Delta$ on the fidelity of the TQD method are dependent on each other and even the fidelity of the TQD method is apparently dependent on the value of $\Delta/t_f$ as mentioned in the last subsection. Significantly, we notice that the fidelities of the two methods are both over 0.98 even when $|\delta x/x|=0.1~(x=t_f,\varepsilon,\Delta)$. Therefore, both of our schemes are robust against the variations in the parameters.

\begin{figure}[htb]\centering
  % Requires \usepackage{graphicx}
  \includegraphics[scale=0.75]{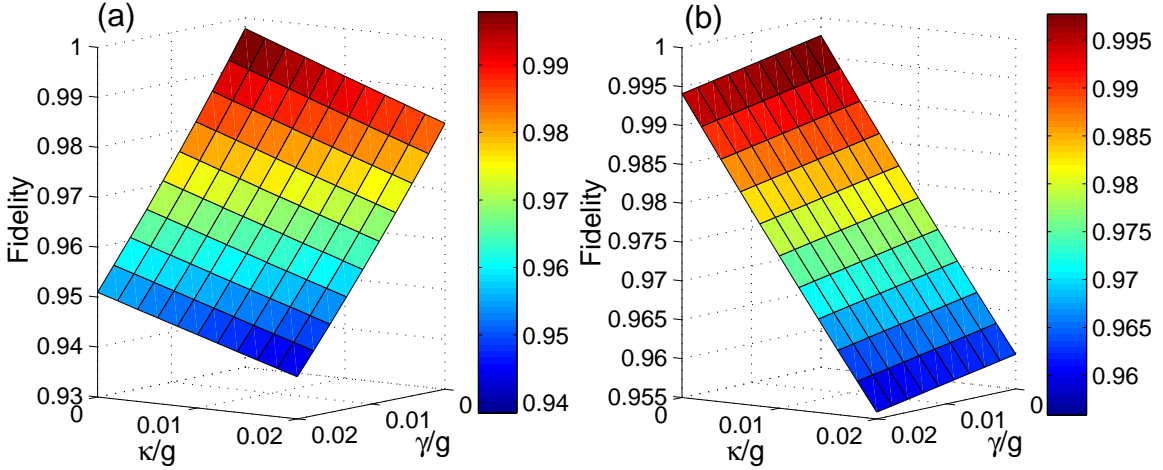}\\
\caption{(Color online) The fidelity of \textbf{(a)}~the LRI method and \textbf{(b)}~the TQD method for generating tree-type three-dimensional entanglement versus $\kappa/g$ and $\gamma/g$, respectively. The parameters used here are same as in Fig.~\ref{f6}.}\label{f9}
\end{figure}
We can also see that the smallest fidelity of the TQD method in Fig.~\ref{f8}~\textbf{(b)} is slightly higher than the smallest fidelity of the LRI method in Fig.~\ref{f8}~\textbf{(a)}. This fact can easily be found by comparing Fig.~\ref{f3}\textbf{(c)} and Fig.~\ref{f5}\textbf{(c)}, in which there is a greater advisable range of the parameters in the TQD method than in the LRI method for preparing tree-type three-dimensional entanglement with a high fidelity.

Next taking the decoherence induced by the atomic spontaneous emissions and photon leakages of the cavity-fiber system into account, the whole system is dominated by the master equation
\begin{eqnarray}\label{e999}
\dot{\rho}(t)&=&-i[H,\rho(t)]\nonumber\\
\cr&&-\sum_{i=L,R}\sum_{j=1,3}\frac{\gamma_{i}^{j}}{2}\Big(\sigma_{e_i,e_i}^{j}\rho-2\sigma_{g_i,e_i}^{j}\rho\sigma_{e_i,g_i}^{j}+\rho\sigma_{e_i,e_i}^{j}\Big)\nonumber\\
\cr&&-\sum_{i=L,R}\sum_{j=1,3}\frac{\gamma_{i}^{j}}{2}\Big(\sigma_{e_i,e_i}^{j}\rho-2\sigma_{g_0,e_i}^{j}\rho\sigma_{e_i,g_0}^{j}+\rho\sigma_{e_i,e_i}^{j}\Big)\nonumber\\
\cr&&-\sum_{i=L,0,R}\frac{\gamma_{i}^{2}}{2}\Big(\sigma_{e_0,e_0}^{2}\rho-2\sigma_{g_i,e_0}^{2}\rho\sigma_{e_0,g_i}^{2}+\rho\sigma_{e_0,e_0}^{2}\Big)\nonumber\\
\cr&&-\sum_{i=L,R}\sum_{j=1,2}\frac{\kappa_{j,i}^f}{2}\Big(b_{j,i}^{\dag}b_{j,i}\rho-2b_{j,i}\rho b_{j,i}^{\dag}+\rho b_{j,i}^{\dag}b_{j,i}\Big)\nonumber\\
\cr&&-\sum_{i=L,R}\sum_{j=1,2,3}\frac{\kappa_{j,i}^c}{2}\Big(a_{j,i}^{\dag}a_{j,i}\rho-2a_{j,i}\rho a_{j,i}^{\dag}+\rho a_{j,i}^{\dag}a_{j,i}\Big),
\end{eqnarray}
where $H$ is the total Hamiltonian $H_{\rm total}$ of the LRI method in Eq.~(\ref{e10}) or $H'_{\rm total}$ of the TQD method in Eq.~(\ref{e31}). $\gamma_{i}^{j}~(i=L,0,R,~j=1,2,3)$ is the spontaneous emission rate of $j$th atom from the excited state $|e_i\rangle_j$ to the ground state $|g_i\rangle_j$; $\kappa_{j,i}^f$ and $\kappa_{j,i}^c$ denote the photon leakage rates
from the fiber modes and the cavity modes , respectively; $\sigma_{mn}^{j}=|m\rangle_{j}\langle n|$ ($m,
n=e_{i},g_{i}$) is Pauli operators. For simplicity, we assume $\gamma_{i}^{j}=\gamma$ and $\kappa_{j,i}^f=\kappa_{j,i}^c=\kappa$.

Based on the master equation, we plot the fidelities of the LRI method and the TQD method versus $\kappa/g$ and $\gamma/g$ in Fig.~\ref{f9}. As we can see from the decrease of the fidelity  of the LRI method with the increases of $\kappa/g$ and $\gamma/g$ in Fig.~\ref{f9}~\textbf{(a)}, we learn that the influence of atomic spontaneous emissions on the fidelity is greater than that of photon leakages of the cavity-fiber system. However, in Fig.~\ref{f9}~\textbf{(a)} the influence of cavity-fiber photon leakages on the fidelity of the TQD method plays a full role, but that of atomic spontaneous emissions is little. As a cross reference, we can get some inspiration from Fig.~\ref{f7}~\textbf{(a)}. In Fig.~\ref{f7}~\textbf{(a)}, the highest value of the cavity-fiber excited populations~(blue dashed line) of the TQD method is over 0.2 during the evolution process but that of the atomic excited populations~(blue solid line) of the TQD method is near zero which caused by the detuning $\Delta$. The highest values of the atomic excited populations~(red solid line) and the cavity-fiber excited populations~(red dashed line) of the LRI method are slightly higher and slightly lower than 0.05, respectively. Therefore, it is no doubt that the results of Fig.~\ref{f7} and Fig.~\ref{f9} are corresponding to each other. Finally, it is necessary to emphasize that the fidelities of the LRI method and the TQD method are near 0.94 and over 0.955 respectively, even when $\kappa=\gamma=0.02g$. Therefore, our two schemes of the LRI method and the TQD method both are robust against the decoherence induced by the atomic spontaneous emissions and photon leakages of the cavity-fiber system.

\section{Experimental feasibility and conclusion}
\label{sec6}
Now we show the experimental feasibility of our schemes. As mentioned in the reference~\cite{SSW2016}, $^{87}\emph{Rb}$ can be used in our schemes to construct the required atomic level configurations. For $^{87}\emph{Rb}$, $|F=1,m_{F}=-1\rangle$, $|F=1,m_{F}=0\rangle$ and $|F=1,m_{F}=1\rangle$ of 5$S_{1/2}$ can be used as the ground states $|g_L\rangle$, $|g_{0}\rangle$ and $|g_R\rangle$ respectively, and $|F=1,m_{F}=-1\rangle$, $|F=1,m_{F}=0\rangle$ and $|F=1,m_{F}=1\rangle$ of 5$P_{3/2}$ can be used as the excited states $|e_L\rangle$, $|e_{0}\rangle$ and $|e_R\rangle$ respectively. As is achieved in recent experiments~\cite{SKV2005,HBP2006,BDR2007} with a set of cavity QED parameters $g=2\pi\times750$MHz, $\gamma=2\pi\times3.5$MHz and $\kappa=2\pi\times2.62$MHz, we will obtain the very high fidelities $F_{LRI}=0.984$ and $F_{TQD}=0.990$ corresponding to the LRI method and the TQD method respectively, which show our schemes to prepare tree-type three-dimensional entangled states both are feasible in the experiment.

In conclusion, we have proposed two schemes to speed up the generations of the tree-type three-dimensional entanglement via Lewis-Riesenfeld invariants and transitionless quantum driving. The two tree-type three-dimensional entangled states are prepared among three atoms trapped respectively in three spatially separated optical cavities which are connected by two fibers. The operation time $t_f=80/g$ is far shorter than $t=3000/g$ which is the generation time of the tree-type three-dimensional entanglement generated in the reference~\cite{SSW2016}. The strict numerical simulations show that the LRI method and the TQD method both are feasible and robust against the variations in the parameters, atomic spontaneous emissions and photon leakages of the cavity-fiber system. Besides, comparing the two methods, we know they both have their own advantages and disadvantages. So we can choose different methods depending on different conditions in experiment. In short, both of our schemes are fast, feasible and robust. We hope that tree-type three-dimensional entanglement will contribute to the improvement of quantum communication security and our work will be useful for the experimental realization of quantum information in the near future.

\begin{center}
{\bf{ACKNOWLEDGMENT}}
\end{center}
This work was supported by the National Natural Science Foundation of China under Grants No. 11464046 and No. 61465013.

\end{document}